\documentclass[aps,pra,showpacs,twocolumn]{revtex4-1}

\usepackage[english]{babel}
\usepackage[utf8]{inputenc}
\usepackage{graphicx}
\usepackage{epstopdf}
\usepackage{amsmath, amssymb}
\usepackage{dcolumn}
\usepackage{subfigure}

\begin{document}

\title{\bf Decoherence of a uniformly accelerated finite-time detector}

\author{Helder A. S. Costa }\email[hascosta@ufpi.edu.br]{ }

\affiliation{ Departamento de F\'{\i}sica, Universidade Federal do Piau\'{\i}, 64049-550 Teresina, PI, Brazil}

\begin{abstract}
 
 We study a uniformly accelerated detector coupled to a massless scalar field for a finite time interval. By considering the detector initially prepared in a superposition state, qubit state, we find that the acceleration induces decoherence on the qubit. Our results suggest the dependence of loss of coherence on the polar angle of qubit state on a Bloch sphere and the time interaction. The adjust those parameters can  significantly improve the conditions to estimate the degree of decoherence induced by Unruh radiation.

\pacs{03.67.Mn, 03.65.Ud, 04.62.+v}
\end{abstract}

\maketitle

 \section{ Introduction}
 
 Accordingly to Unruh and Wald \cite{Unruh} a uniformly accelerated detector (i.e., two-level atom) coupled to a massless scalar field in the Minkowski vacuum perceive a thermal distribution of Rindler particles with a temperature proportional to its proper acceleration. This effect is usually named as Fulling-Davies-Unruh effect or simply Unruh effect \cite{Unruh2}. In original proposal, the Unruh effect has been analyzed in an in-out approach, i.e., the initial state of the system - detector plus quantum scalar field - is assumed to be prepared at past infinity and the out state is evaluated at infinite future. On the other hand, in \cite{Svaiter, Higuchi, Padmanabhan} it was investigated the response of uniformly accelerated detectors when they are coupled to the quantum scalar field for a finite period of time. In particular, it was verified that the response of a finite-time detector depend on the manner in which it is switched on and off, for example, the response is divergent for abrupt switching of the detector. In order to avoid divergences, the authors point out that the detector must be switched on and off via a smooth window function (e.g., Gaussian function).    

 In all these works, the composite system consisting of the detector plus the quantum scalar field were describe as a closed system. In this paper, our purpose is study the effect of decoherence induced by acceleration on a finite-time detector prepared in a superposition state. Thus, we assume that the uniformly accelerated detector behaves like an open system and the vacuum with the fluctuations of the field as its environment. In this context, if one only observes the state of the detector and not the field, then the detector will be found in either the excited state or the ground state. Consequently, the detector state will no longer be a pure state, but the superposition state of detector becomes a statistical mixture after their interaction with the scalar field. This represents a physical process of loss of coherence. It is worth mentioning that in this process the coherence is thereby not destroyed but delocalized into the degrees of freedom of the field where it is physically inaccessible.      

 The paper is organized as follows. In Sec. II, we introduce the standard Unruh-DeWitt detector model that characterize the couple between the detector (two-level system) and quantum scalar field. Also, we evaluate finite-time corrections to the rate of transition probability. In Sec. III, we present our scheme where a uniformly accelerated detector prepared in a superposition state interacts with a scalar field for a finite time interval and then their internal states are measured. In Sec. IV, we quantify the degree of decoherence via fidelity measure. Finally, our concluding remarks are addressed in Sec. V. 


\section{The model} 
In the interaction picture, the detector with internal levels $|g\rangle$ and $|e\rangle$ is coupled to a massless scalar field $\phi$ according to the interaction Hamiltonian $\hat{\mathcal{H}}_{int}$ in the detector frame
\begin{align} \label{Hint}
\hat{\mathcal{H}}_{int} = \hbar g\Theta(\tau)\phi[x(\tau)][\hat{\sigma}_{+}e^{i\Omega\tau} + \hat{\sigma}_{-}e^{-i\Omega\tau}],
\end{align}
where $\Omega$ is the transition angular frequency and $\tau$ is
the atom proper time. $\hat{\sigma}_{+} = |e\rangle\langle g|$ and
$\hat{\sigma}_{-} = |g\rangle\langle e|$ are the raising and
lowering operators, respectively, and $g$ is the
effective coupling constant between the detector and the field. 
In order to describe finite-time interaction, we assume that $\Theta(\tau)$ is a gradual window function with the properties: $\Theta(\tau) \approx 1$ for $|\tau| \ll T$ and  $\Theta(\tau) \approx 0$ for $|\tau| \gg T$ where $T$ is the interaction time. It is worthy to mention that the
Hamiltonian, given by Eq. (\ref{Hint}), describes the atom-field
interaction in the dipole approximation.

  Consider that the detector is in motion on a trajectory $x(\tau)$ and the field is in the Minkowski vacuum state $|0_{\mathcal{M}}\rangle$, the probability of excitation of the detector with simultaneous photon emission, up to the first order in perturbation theory, is given by
\begin{align*} 
\mathcal{P} = \int_{-\infty}^{\infty} d\tau  \int_{-\infty}^{\infty} d\tau' \Theta(\tau)\Theta(\tau') e^{-i\Omega(\tau - \tau')} G^{+}(x(\tau), x(\tau')) 
\end{align*}
where $G^{+}(x(\tau), x(\tau')) = \langle 0_{\mathcal{M}}|\phi[x(\tau)]\phi[x(\tau')]|0_{\mathcal{M}}\rangle$ is the Wightman function \cite{Birrell}.  For inertial and accelerated trajectories in Minkowski space, the Wightman function is invariant under time translation in the detector frame, i.e., $G^{+}(x(\tau), x(\tau')) = G^{+}(\tau - \tau')$ \cite{Padmanabhan2, Letaw}. By using the following property $f(u)[e^{-i\Omega u}G^{+}(u)] = f(-i\frac{\partial}{\partial \Omega})[e^{-i\Omega u}G^{+}(u)]$ for any function $f(u)$ which has a power series expansion around $u = 0$, we can provide an asymptotic expression for the probability of transition with any smooth window function,
  \begin{align*}
 \mathcal{P} &= \Theta\left(i\frac{\partial}{\partial\Omega}\right)\Theta\left(-i\frac{\partial}{\partial\Omega}\right)\mathcal{P}^{\infty}, 
 \end{align*}
where $\mathcal{P}^{\infty}$ correspond to the infinite-time detector
\begin{align*}
\mathcal{P}^{\infty} &= \int_{-\infty}^{\infty} d\tau  \int_{-\infty}^{\infty} d\tau' e^{-i\Omega(\tau - \tau')}G^{+}(\tau - \tau').
\end{align*}
Expanding $\Theta(\tau)$ as a Taylor series around $\tau = 0$ and assuming that $\Theta(0) = 1$ and $\Theta'(0) = 0$, we obtain that $\mathcal{P} \approx \mathcal{P}^{\infty} - \Theta''(0)\frac{\partial^2\mathcal{P}^{\infty}}{\partial\Omega^2}$. The corresponding transition probability per unit time is given by $\mathcal{R} \approx \mathcal{R}^{\infty} - \Theta''(0)\frac{\partial^2\mathcal{R}^{\infty}}{\partial\Omega^2}$ where $\mathcal{R}^{\infty} = \int_{-\infty}^{\infty} d\Delta\tau e^{-i\Omega\Delta\tau}G^{+}(\Delta\tau)$ with $\Delta\tau = \tau - \tau'$. Notice that the rate of transition probability depends on the derivatives of the window function. Hence, if the detector is turned on and off abruptly these derivatives leading to divergent responses. In order to avoid divergences, let us assume that the detector is switched on and off by the Gaussian window function $ \Theta(\tau) = \exp\left(-\frac{\tau^2}{2T^2}\right)$. In this case, to the leading order, the finite-time corrections to the rate of transition probability reads
 \begin{align} \label{R}
 \mathcal{R} \approx \mathcal{R}^{\infty} + \frac{1}{2T^2}\frac{\partial^2\mathcal{R}^{\infty}}{\partial\Omega^2} + \mathcal{O}\left(\frac{1}{T^4}\right).
\end{align}
In particular, for a uniformly accelerated detector moving with proper acceleration $a$, the natural coordinate system in detector frame is related to the Minkowski coordinates by the transformations \cite{Rindler}:
\begin{align*} 
t(\tau) = \frac{c}{a}\sinh\left(\frac{a\tau}{c}\right),\quad z(\tau) =
\frac{c^2}{a}\cosh\left(\frac{a\tau}{c}\right).
\end{align*} 
 In terms of this coordinates, the Wightman function can be expressed as $G^{+}(\Delta\tau) = -\frac{1}{4\pi^2}\sum_{n=-\infty}^{\infty}\frac{1}{(\Delta\tau - 2i\epsilon + 2\pi in/a)^2}$ and the transition probability rate $\mathcal{R}^{\infty}$ becomes
 \begin{align} \label{Racc}
 \mathcal{R}^{\infty}_{\mathrm{acc}} = \frac{1}{2\pi}\frac{\Omega}{e^{\frac{2\pi\Omega c}{a}} - 1},
 \end{align}
 which is the standard thermal spectrum. This result shows that a uniformly accelerated detector in the Minkowski vacuum responds as it would if it were at rest in a thermal bath with a temperature that is propotional to the acceleration $T_{\mathrm{U}} = \frac{\hbar a}{2\pi k_{B} c}$ (Unruh temperature). Here $\hbar$ begin the Planck's constant, $k_B$ the Boltzmann's constant and $c$ the light speed in vacuum. By substituting (\ref{Racc}) in (\ref{R}), we find

\begin{align} \label{A3}
 \mathcal{R}_{\mathrm{acc}} &\approx \frac{1}{2\pi}\frac{\Omega}{e^{\frac{2\pi\Omega c}{a}} - 1}
 \Bigg\lbrace 1 + \frac{2\pi c}{a\Omega T^2} \frac{e^{\frac{2\pi\Omega c}{a}}}{e^{\frac{2\pi\Omega c}{a}} - 1}  \nonumber \\
&\times  \left[1 - e^{\frac{2\pi\Omega c}{a}} + \frac{\pi\Omega c}{a}\left(e^{\frac{2\pi\Omega c}{a}} + 1\right) \right] \Bigg\rbrace 
\end{align}
 This is an approximate expression for the rate of transition probability when the detector is gradually coupled to a scalar field during a finite time interval $T$. It is clear from Eq. (\ref{A3}) that in the limit $T \rightarrow \infty$, the first term of expression above dominates, and we recover the standard result.


\section{The setup} 
 
\begin{figure}[h]
\centering
\includegraphics[width=0.47\textwidth]{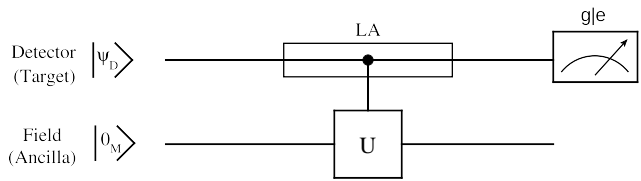}
\caption{Illustration of a quantum network for estimate the loss of coherence of a single qubit induced by Unruh radiation. The detector, initially in the superposition state  $|\psi_{\mathrm{D}}\rangle$, is uniformly accelerated in LA, interacts with a scalar field for a finite time interval $T$, and then their internal states are measured. The U operator models the interaction between the qubit and an auxiliary field.} \label{fig1}
\end{figure}

 The basic idea of our scheme is as follows. As shown in Fig. \ref{fig1},  we consider the setting that consists of a detector and an auxiliary quantum field. The field is in the vacuum state $|0_{\mathcal{M}}\rangle$, while the detector is prepared in the superposition state $|\psi_{\mathrm{D}}\rangle = \alpha|g\rangle + \beta|e\rangle$ with $|\alpha|^2 + |\beta|^2 = 1$, i.e., the detectors are in a qubit state. In this case, it is convenient to rewrite $|\psi_{\mathrm{D}}\rangle$ as a Bloch vector 
\begin{align}
|\psi_{\mathrm{D}}\rangle = \cos\frac{\theta}{2}e^{i\chi/2}|g\rangle + \sin\frac{\theta}{2}e^{-i\chi/2}|e\rangle
\end{align}
where $\theta \in [0,\pi]$ and $\chi \in [0,2\pi]$ are the polar and azimuthal angles in the Bloch sphere, respectively. After its preparation, the detector is uniformly accelerated in a linear accelerator (LA), 
interacts with the scalar field $\phi$ for a finite time interval $T$, and finally its internal states are measured.

 Let us analyze in more detail the scheme shown in Fig. 1, suppose that initially the system (accelerated detector plus scalar field) is prepared in the state 
$|\psi_\mathrm{in}\rangle \rightarrow |0_{\mathcal{M}}\rangle\otimes|\psi_{\mathrm{D}}\rangle,$
where $|0_{\mathcal{M}}\rangle$ is the Minkowski vacuum state of the field
defined in the inertial laboratory frame. Considering that the interaction between the detector and the field is governed by (\ref{Hint}). In the weak coupling regime, the unitary transformation induced by the Hamiltonian (\ref{Hint}) is given by
\begin{align*}
  \hat{U} = I - ig\int_{-\infty}^{\infty}d\tau\Theta(\tau)\phi[x(\tau)][\hat{\sigma}_{+}e^{i\Omega\tau} + \hat{\sigma}_{-}e^{-i\Omega\tau}].
\end{align*}
From above result, we can obtain that the interaction between the uniformly accelerated detector and the field produces the following transformation
\begin{align}\label{atomfield}
|0_{\mathcal{M}}\rangle\otimes|g\rangle &\rightarrow |0_{\mathcal{M}}\rangle\otimes|g\rangle -
ig\Phi(\tau)|0_{\mathcal{M}}\rangle\otimes|e\rangle,\nonumber\\
|0_{\mathcal{M}}\rangle\otimes|e\rangle &\rightarrow |0_{\mathcal{M}}\rangle\otimes|e\rangle -
ig\Phi^{*}(\tau)|0_{\mathcal{M}}\rangle\otimes|g\rangle. 
\end{align}
 where $\Phi(\tau) = \int_{-\infty}^{\infty}d\tau\Theta(\tau)e^{i\Omega\tau}\phi[x(\tau)]$. Note that (\ref{atomfield}) describes the ``dressing" of the ground and excited states of the uniformly accelerated detector by the interaction.
 Here we are interested in observe only the effect of the acceleration radiation. Thus, we consider that the lifetime of the excited state is greater than the interaction time. This avoids that the atom making a transition to the ground state by spontaneously emitting one photon into the field \cite{Scully, Schleichbook}.   
Therefore, from Eq. (\ref{atomfield}), after of the finite-time interaction between the accelerated detector and the field, the state of the system reads as
\begin{align*}
|\psi_{\mathrm{out}}\rangle &\rightarrow |0_{\mathcal{M}}\rangle\otimes|\psi_{\mathrm{D}}\rangle  - ige^{i\chi/2}\cos\frac{\theta}{2} \Phi(\tau)|0_{\mathcal{M}}\rangle\otimes|e\rangle \\
& - ige^{-i\chi/2}\sin\frac{\theta}{2} \Phi^{*}(\tau)|0_{\mathcal{M}}\rangle\otimes|g\rangle  .
\end{align*}
 If one only observes the state of the detector and not the field, then the detector will be found in either the excited state or the ground state; nevertheless, it will no longer be in a pure state. The new state can be described by reduced density matrix $\hat{\rho}_{\mathrm{D}} = \mathrm{Tr}_{\mathcal{M}}[|\psi_{\mathrm{out}}\rangle\langle\psi_{\mathrm{out}}|]$. In the basis $\{|g\rangle, |e\rangle\}$, the reduced density matrix of the detector is given by
\begin{align*}
&\hat{\rho}_{\mathrm{D}} = \frac{1}{1 + g^2T\mathcal{R}_{\mathrm{acc}}} \\
&\times\left(\begin{array}{cc}
 \cos^2\frac{\theta}{2} + g^2T\mathcal{R}_{\mathrm{acc}}\sin^2\frac{\theta}{2} & e^{-i\chi/2}\sin\frac{\theta}{2}\cos\frac{\theta}{2} \\
 e^{i\chi/2}\sin\frac{\theta}{2}\cos\frac{\theta}{2} &  \sin^2\frac{\theta}{2} + g^2T\mathcal{R}_{\mathrm{acc}}\cos^2\frac{\theta}{2}
\end{array} \right).
\end{align*}
where $\mathcal{R}_{\mathrm{acc}}$ is given in (\ref{A3}). Hence the accelerated qubit is expected to exhibit a loss of coherence, in which the pure state $|\psi_{\mathrm{D}}\rangle$ reduce to a statistical mixture after their interaction with the scalar field. Note that this loss of coherence takes some time and its duration is named as decoherence time $\tau_{\mathrm{d}}$ \cite{Audretsch}. From the uncertainly relation $\Delta\tau\Omega \gtrsim 1$ with $\Delta\tau$ being the scale time for transition between the two levels of detector, one sees that the interaction cannot be performed during a time interval less than $\frac{1}{\Omega}$. This gives a lower limit for the decoherence time and it has to be included into our theoretical approach for the detector to work effectively.

 \section{Quantifying decoherence} 
 
\begin{figure}[h]
\centering
\includegraphics[width=0.45\textwidth]{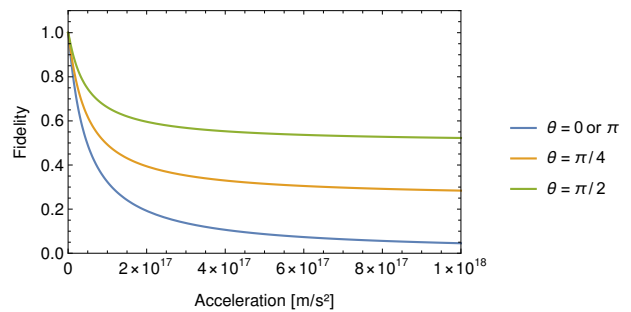}
\caption{Fidelity as a function of the acceleration for different values of parameter $\theta$. Here we have fixed $T = 1$ $\mu$s, $g = 0.5$ Hz and $\Omega = 1$ MHz. }
\label{fig2}
\end{figure}

   One way to quantify the degree of decoherence induced by acceleration on the qubit is  evaluating the  state fidelity between the initial state $|\psi_{\mathrm{D}}\rangle$ and the final state $\hat{\rho}_{\mathrm{D}}$, i.e., $\mathcal{F} = \mathrm{Tr}[|\psi_{\mathrm{D}}\rangle\langle\psi_{\mathrm{D}}|\hat{\rho}_{\mathrm{D}}]$, which is calculated to be
\begin{align}
\mathcal{F} = \frac{1 + 2g^2T\mathcal{R}_{\mathrm{acc}}\cos^2\frac{\theta}{2}\sin^2\frac{\theta}{2}}{1 + g^2T\mathcal{R}_{\mathrm{acc}}}.
\end{align}
The Fig. \ref{fig2} shows the dependence of the fidelity $\mathcal{F}$ as a function of acceleration for different input state conditions. In all cases, the plots show loss of coherence due to acceleration. In particular, note that for $\theta = \pi/2$ the fidelity decreases and approaches its asymptotic value $1/2$ with the increase of acceleration, which indicates a partial loss of coherence. The angle $\theta = \pi/2$ indicate input states $|\psi_{\mathrm{D}}\rangle = \frac{1}{\sqrt{2}}[|g\rangle \pm |e\rangle]$ with $\chi = 0,\pi$. On the other hand, when $\theta = 0, \pi$ the fidelity decrease to its asymptotic minimum $\mathcal{F} \rightarrow 0$, indicating a higher decoherence process. Notice that the angles $\theta = 0, \pi$ represent states $|g\rangle$ and $|e\rangle$, i.e., eigenstates of $\sigma_{\mathrm{z}}$. This results suggests a dependence of degree of decoherence induced by acceleration on the input state conditions.

\begin{figure}[h]
\centering
\includegraphics[width=0.45\textwidth]{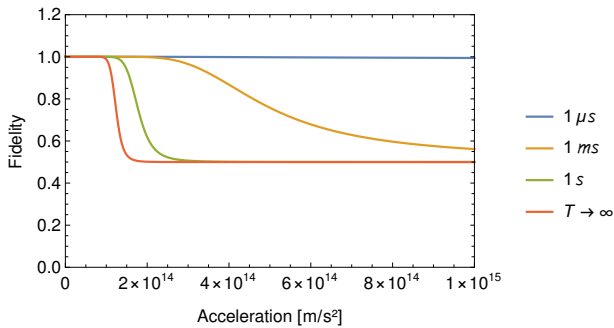}
\caption{Fidelity as a function of the acceleration for different values of interaction time $T$. Here we have fixed $\theta = \pi/2$, $g = 0.5$ Hz and $\Omega = 1$ MHz. }
\label{fig3}
\end{figure}

We are also interested in how the interaction time $T$ influence the loss of coherence due to acceleration. In Fig. \ref{fig3}, we plot the fidelity as a function of the acceleration for different values of interaction time. Notice that, when $T = 1$ $\mu$s the fidelity slowly decreases when the acceleration grows. On the other hand, we can observe that the fidelity decreases notably and approaches its asymptotic value $1/2$  for a long interaction time ($T \rightarrow \infty$). This asymptotic regime corresponds to when $T \gg \frac{1}{\Omega}$, i.e., when the time interaction is very larger compared with time scale for the transition in the detector. Also note that for long interaction time the degree of decoherence can be observable for accelerations as low as $10^{14}\;\mathrm{m/s^2}$. 
 
 It is interesting to discuss our results in connection with the previous work \cite{Kok, Li, Nesterov}. In \cite{Kok}, Kok and Yurtsever investigate an accelerating qubit in vacuum of Minkowski. By measuring the purity of accelerated detector, it was found that the qubit decoheres due to thermal Unruh radiation. Our scheme is physically different and represents a step forward in several aspects. (1) We consider a uniformly accelerated detector coupled to the scalar field for a finite time interval. This correspond to a more realistic situation. (2) We have analyzed the loss of coherence as a function of both the detector initial preparations (i.e., the polar angle of qubit state on a Bloch sphere) and the time interaction. (3) We show that the adjust those parameters can significantly improve the conditions to estimate the decoherence induced by Unruh radiation. In addition, our results suggest that the loss of coherence is significative for relatively low accelerations. 

From a practical point of view several conditions must be considered with respect to the feasibility of our scheme: (1) The detector should be considered a stable two-level atom with the lifetime of the excited state greater than the interaction time. A good example is the Helium atom (He*) with electric dipole transition ($2 ^3\mathrm{S}_1 \rightarrow 2 ^3\mathrm{P}$) and natural linewidth of $1.6\;\mathrm{MHz}$. In addition, the timelife of the Helium atom is of the order of $8 \times 10^3\;\mathrm{s}$ which can be considered to be stable for the present application \cite{Vassen}. (2) The acceleration mechanism must  be performed so that it does not ionize or strongly perturb the internal states of the detector as suggested in \cite{Martinez}. In particular, short-pulse lasers have been used in acceleration of Rydberg-like atoms in \cite{Eichmann, McWilliams}. (3) The coupling  between the detector and the field can be perform by using a microwave cavity with perfect mirrors. (4) In an experiment of finite time interaction the detector will be affected by the transients related to the temporal change of the window function $\Theta(\tau)$. (5) As discussed in the previous section, one has to impose the condition $\Delta\tau\Omega \gtrsim 1$ in order to keep the disturbance introduced by the vacuum fluctuations on the detector sufficiently small. Under this condition, the detector can work effectively due to an appropriate decoherence time $\tau_{d}$.

\section{Conclusions} 

  In summary, we present the model of a uniformly accelerated detector interacting with a quantum scalar field for a finite time interval. In particular, we have analyzed the loss of coherence induced by acceleration. This decoherence is quantifying via the reduction of the fidelity. Compared to previous studies about accelerating qubit in vacuum of Minkowski, our study represents a step forward to provide a quantitative analysis of the loss of coherence as a function of both the detector initial preparations and the time interaction. We arrive at the conclusion that the adjust those parameters can provide us a better estimation of the decoherence induced by Unruh effect. Thus, our results provide a novel insight for efficient experimental strategies in the estimation of Unruh effect via decoherence process.






\section*{Acknowledgments} 

 This work was supported by the Brazilian funding agency CAPES.


\begin{thebibliography}{99}

\bibitem{Unruh} W. G. Unurh and R. M. Wald, Phys. Rev. D \textbf{29}, 1047 (1984).


\bibitem{Unruh2} W. G. Unruh, Phys. Rev. D \textbf{14}, 870 (1976); S. A. Fulling, Phys. Rev. D \textbf{7}, 2850 (1973); P. Davies, J. Phys. A \textbf{8}, 609 (1975).



\bibitem{Svaiter} B. F. Svaiter and N. F. Svaiter, Phys. Rev. D \textbf{46}, 5267 (1992).

\bibitem{Higuchi} A. Higuchi, G. E. A. Matsas, and C. B. Peres, Phys. Rev. D \textbf{48}, 3731 (1993). 

\bibitem{Padmanabhan} L. Sriramkumar and T. Padmanabhan, Class. Quantum Grav. \textbf{13}, 2061 (1996).


\bibitem{Birrell} N. D. Birrell and P. C. W. Davies, Quantum Fields InCurved Space, Cambridge monographs on mathematical physics (Cambridge University Press, Cambridge, 1982).

\bibitem{Padmanabhan2} T. Padmanabhan, Astrophys. Space. Sci. \textbf{83}, 247 (1982).

\bibitem{Letaw} J. R. Letaw and J. D. Pfautsch, Phys. Rev. D \textbf{24}, 1491 (1982).


\bibitem{Rindler}  W. Rindler, Am. J. Phys. \textbf{34}, 1174 (1966).




\bibitem{Scully} M. O. Scully {\it et al.}, Phys. Rev. Lett. \textbf{91}, 243004 (2003).

\bibitem{Schleichbook} W. P. Schleich, Quantum Optics in Phase Space (Berlin: Wiley-VCH, 2001).

\bibitem{Audretsch} J. Audretsch, M. Mensky, and R. Müller, Phys. Rev. D \textbf{51}, 1716 (1995).



\bibitem{Kok} P. Kok and U. Yurtsever, Phys. Rev. D \textbf{68}, 085006 (2003).

\bibitem{Li} Y. Li, Y. Dai, and Y. Shi, Eur. Phys. J. C \textbf{77}, 598 (2017).

\bibitem{Nesterov} A. I. Nesterov, G. P. Berman, M. A. R. Fernández, and X. Wang, arXiv:2003.05014 [gr-qc].



\bibitem{Vassen} W. Vassen, R. P. M. J. W. Notermans, R. J. Rengelink, and R. F. H. J. v. d. Beek, Appl. Phys. B \textbf{122}, 289 (2016).

\bibitem{Martinez} E. Martín-Martínez, I. Fuentes, and R. B. Mann, Phys. Rev. Lett. \textbf{107}, 131301 (2011).


\bibitem{Eichmann} U. Eichmann, T. Nubbemeyer, H. Rottke, and W. Sandner, Nature \textbf{461}, 1261 (2009).

\bibitem{McWilliams} C. Maher-McWilliams, P. Douglas, and P. F. Barker, Nature Photon. \textbf{6}, 386 (2012).
























\end{thebibliography}
\end{document}